\documentclass[aps,pra,twocolumn,showpacs,superscriptaddress,groupedaddress]{revtex4}  
\usepackage{graphicx}  
\usepackage{dcolumn}   
\usepackage{bm}        % for math
\usepackage{amssymb}   % for math
\usepackage{caption}
\usepackage{graphicx}
\usepackage[justification=centering]{caption} 
\begin{document}
\captionsetup[figure]{name={Fig.}}
\preprint{APS/123-QED}
\title{Wireless Multihop Quantum Teleportation Utilizing a 4-Qubit Cluster State}
\author{S. J. Emem-Obong}
\affiliation{Department of Physics, Federal University of Lafia, Nigeria}
\author{Yame Mwanzang Philemon}
\affiliation{Department of Physics, Federal University of Lafia, Nigeria}
\author{C. Iyen}
\affiliation{Department of Physics, Federal University of Lafia, Nigeria}
\affiliation{Department of Pure and Applied Physics, Federal University Wukari, Nigeria.}
\author{M. S. Liman}
\affiliation{Department of Physics, Federal University of Lafia, Nigeria}
\author{B. J. Falaye}
\email{babatunde.falaye@science.fulafia.edu.ng}
\affiliation{Department of Physics, Federal University of Lafia, Nigeria}
\date{\today}
\begin{abstract}
This paper proposes a quantum routing protocol using multihop teleportation for wireless mesh backbone networks. After analyzing the quantum multihop protocol, a four-qubit cluster state is selected as the quantum channel for the protocol. The quantum channel between intermediate nodes is established through entanglement swapping, utilizing the four-qubit cluster state. Additionally, both classical and quantum routes are created in a distributed manner. We demonstrate that quantum information can be teleported hop-by-hop from the source node to the destination node. Successful quantum teleportation occurs when the sender performs Bell state measurements (BSM), while the receiver introduces auxiliary particles, applies a positive operator-valued measure (POVM), and uses a corresponding unitary transformation to recover the transmitted state. We analyze the success probability of quantum state transfer and find that the optimal success probability is achieved when $\tau_{2|1} = \frac{1}{\sqrt{2}}$. Our numerical results show the susceptibility of $P_{\text{suc}}$ to the number of hops $N$. These findings indicate that multihop teleportation using distributed wireless quantum networks with a four-qubit cluster state is feasible.
\end{abstract}

\maketitle
\section{Introduction}
In the recent years, there has been incessant avidity in studying multi-user quantum communication because it offers opportunity to construct quantum networks. With quantum networks, quantum information between physically separate quantum systems can be transmitted. In fact, it forms a salient component of quantum computing and quantum cryptography systems. It has been discerned that transmission via quantum teleportation and, directly from one node to another are two methods to transmit an unknown quantum state between two nodes \cite{MA1}. However, node-to-node transmission is only practicable for short distance. Also, since quantum systems unavoidably interact with the environment, node-to-node transmission might easily debase the quantum states. On the contrary, in the method of quantum teleportation, the two nodes do not need to transmit quantum state directly since an unknown quantum state is transmitted via quantum entanglement.  Consequently, it is reasonable to employ this method for wireless transmission over any distance.

Recently, a scheme for faithful quantum communication in quantum wireless multihop networks, by performing quantum teleportation between two distant nodes which do not initially share entanglement with each other, was proposed by Wang et al. \cite{MA2}. Xiong et al. \cite{MA3} proposed a quantum communication network model where a mesh backbone network structure was introduced. Some other pertinent reports can be found in refs. (\cite{MA4,MA5} and references therein). Although several relevant results have been obtained along this direction, but most contributions have been based on 2- and 3-qubit Greenberger-Horne-Zeilinger (GHZ) state as quantum channel. In this paper, we propose quantum wireless multihop network with a mesh backbone structure, based on 4-qubit cluster state via the method of quantum teleportation.

The cluster state \cite{MA6}, which is a type of highly entangled state of multiple qubits, is generated in lattices of qubits with Ising type interactions. On the basis of single qubit operation, the cluster state serves as initial resource for a universal computation scheme \cite{MA7}. Cluster state has been realized experimentally in photonic experiments \cite{MA7} and in optical lattices of cold atoms \cite{MA8}. In this paper we select four-qubit cluster state as the entangled resource. When the state of one particle in the entangled pair changes, the state of other particle situated at a distant node changes non-contemporaneously. Thus, entanglement swapping can be applied. Using a classical communication channel, the results of the local measurement can be transmitted from node-to-node in a secure manner.

\section{Wireless mesh network and routing protocol}
A wireless mesh network (WMN) can be described as a mesh network established through the connection of wireless access points which have been installed at the location of each network users. It consists of mesh routers, which are stationary, and mesh client, which are removable. In WMN, there exist quantum wireless channel and classical wireless channel for communications. The classical channel serves the purpose of classical information transmission while the quantum wireless channel exists between neighbor nodes. Quantum information can be transmitted from node-to-node only when quantum route and classical route co-exist. Classical information is transmitted along classical route while quantum information is via the quantum route.
\begin{widetext}
\begin{figure*}[!t]
\centering \includegraphics[height=60mm, width=165mm]{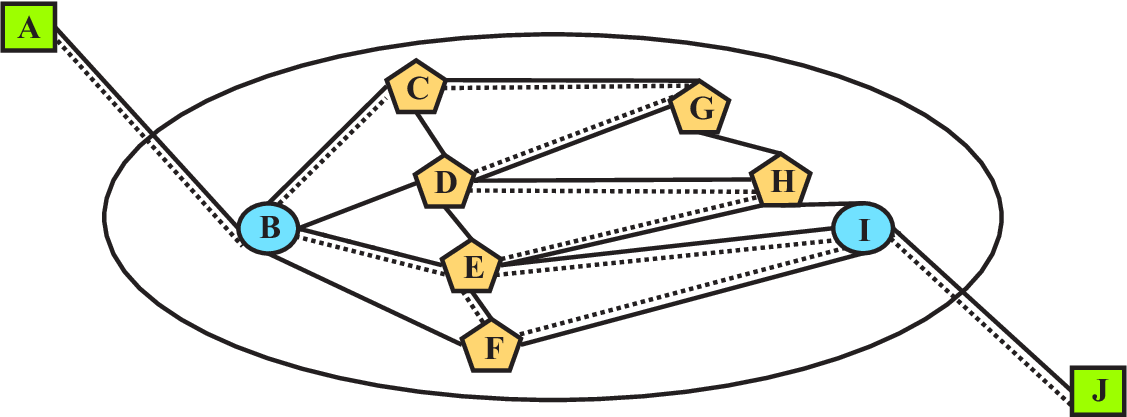}
\caption{\protect\small The quantum mesh backbone network. The dotted lines represent quantum channels while the solid lines denote classical channels. Node A is not directly entangled with the node J. However, quantum channels between them can be established via entanglement swapping.}
\label{figu1}
\end{figure*}
\end{widetext}

In wireless quantum communication, there exists mesh backbone network which consist of route nodes and edge route nodes. We delineate the quantum mesh network in Fig. \ref{figu1}. Node A wishes to send information to node J. To achieve this, it scrutinizes its  routing table to find if there is any available route to J. If there are available route, it forwards the packet to next hop node. However, in the absence of none, source node A requests for a quantum route discovery from the neighboring edge route B and thus, the quantum route finding process commences. Once a routing path that permits co-existence of quantum and classical route, from source node to the destination is found and selected, the edge route node I sends a route reply to node A. At this moment, the process of establishing the quantum channel commences.

\section{Process of establishing the quantum channel}
\begin{figure*}[!t]
\centering \includegraphics[height=140mm, width=165mm]{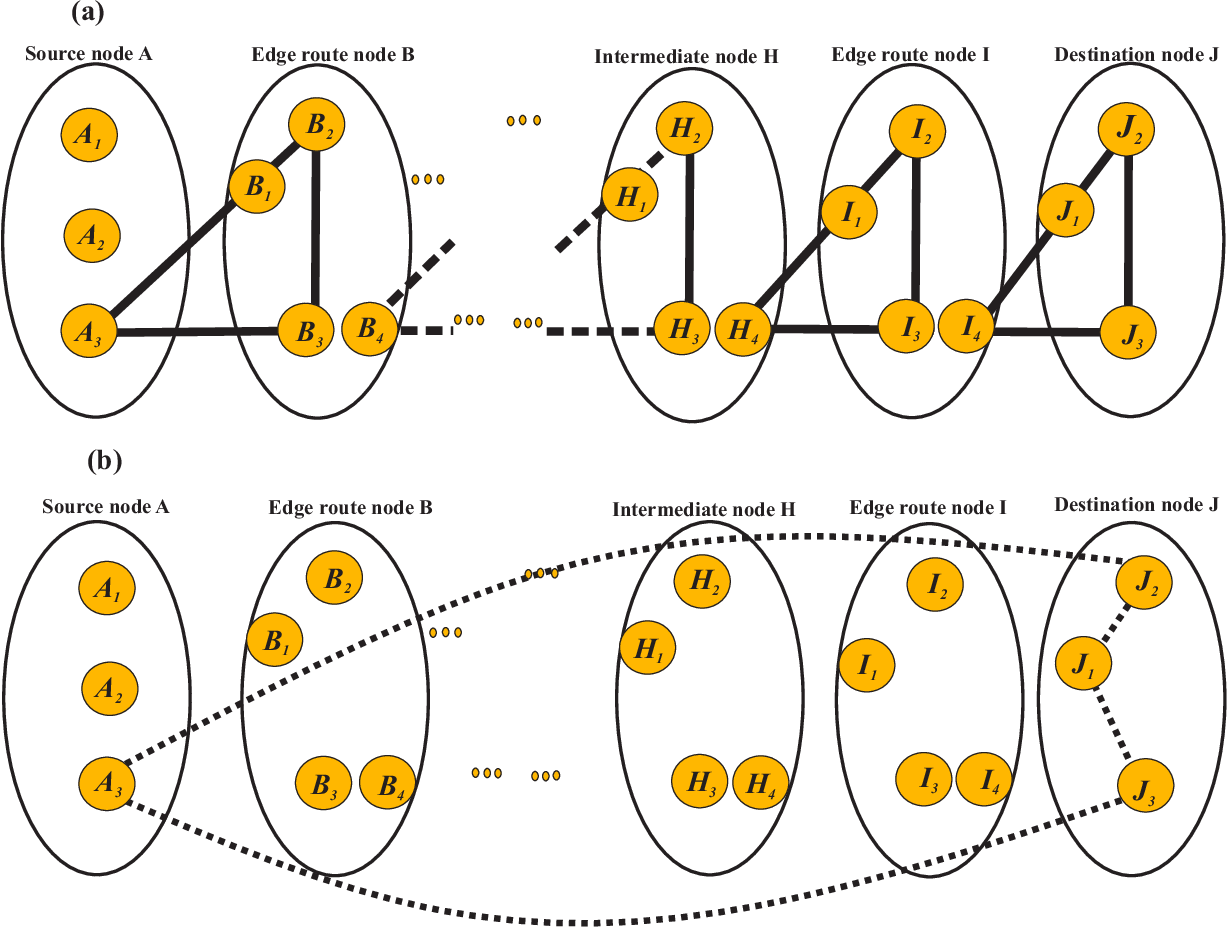}
\caption{\protect\small The process of establishing quantum channel. (a) Before route-finding process. (b) After route-finding process.}
\label{fig2}
\end{figure*}

In this section, we establish the quantum channel linking the nodes. As it can be seen in Figs. \ref{fig2} (a) and (b), A denotes the source node while the destination node is denoted by J. In the source node A, there exists two-qubit of unknown state $\left|\chi\right\rangle_{A_1A_2}=a_0\left|00\right\rangle+d_0\left|11\right\rangle$. Let $N$ denotes number of nodes such that between A and J, we have $N-1$ nodes. The entangle state of the neighboring nodes is 4-qubit cluster state of the form $\left|\mathcal{CS}\right\rangle=1/2\left(\left|0000\right\rangle+\left|0011\right\rangle+\left|1100\right\rangle-\left|1111\right\rangle\right)$. Now, the edge node I performs Bell state measurement on particle pairs $(I_3,I_4)$ and $(I_1,I_2)$ to obtain
\begin{widetext}
\begin{eqnarray}
&&\left|\Pi\right\rangle=\nonumber\\
&&\left|\chi\right\rangle_{A_1A_2}\otimes\frac{1}{8}\sum_{\varsigma,\kappa\in[+,-]}\big[\left|\Phi^\varsigma\right\rangle_{I_3I_4}\left|\Phi^\kappa\right\rangle_{I_1I_2}\left(\left|0000\right\rangle+\left|0011\right\rangle\pm^{(\varrho)}\mp^{(\zeta)}\left|1100\right\rangle\pm^{(\varrho)}\pm^{(\zeta)}\left|1111\right\rangle\right)_{H_4,J_1,J_2,J_3}\nonumber\\
&&\ \ \ \ \ \ \ \ \ \ \ \ \ \ \ \ \ \ +\left|\Psi^\varsigma\right\rangle_{I_3I_4}\left|\Phi^\kappa\right\rangle_{I_1I_2}\left(\left|0100\right\rangle-\left|0111\right\rangle\pm^{(\varrho)}\mp^{(\zeta)}\left|1000\right\rangle\pm^{(\varrho)}\mp^{(\zeta)}\left|1011\right\rangle\right)_{H_4,J_1,J_2,J_3}\nonumber\\
&&\ \ \ \ \ \ \ \ \ \ \ \ \ \ \ \ \ \ +\left|\Phi^\varsigma\right\rangle_{I_3I_4}\left|\Psi^\kappa\right\rangle_{I_1I_2}\left(\pm^{(\varrho)}\left|0100\right\rangle\mp^{(\varrho)}\left|0111\right\rangle\pm^{(\zeta)}\left|1000\right\rangle\pm^{(\zeta)}\left|1011\right\rangle\right)_{H_4,J_1,J_2,J_3}\nonumber\\
&&\ \ \ \ \ \ \ \ \ \ \ \ \ \ \ \ \ \ +\left|\Psi^\varsigma\right\rangle_{I_3I_4}\left|\Psi^\kappa\right\rangle_{I_1I_2}\left(\pm^{(\varrho)}\left|0000\right\rangle\pm^{(\varrho)}\left|0011\right\rangle\pm^{(\zeta)}\left|1100\right\rangle\mp^{(\zeta)}\left|1111\right\rangle\right)_{H_4,J_1,J_2,J_3}\big]\nonumber\\
&&\ \ \ \ \ \ \ \ \ \ \ \ \ \ \ \ \ \ \otimes_{i=3}^N\left|\mathcal{CS}\right\rangle^{n_i},
\label{}
\end{eqnarray}
\end{widetext}
\begin{figure*}[!t]
\centering \includegraphics[height=140mm, width=160mm]{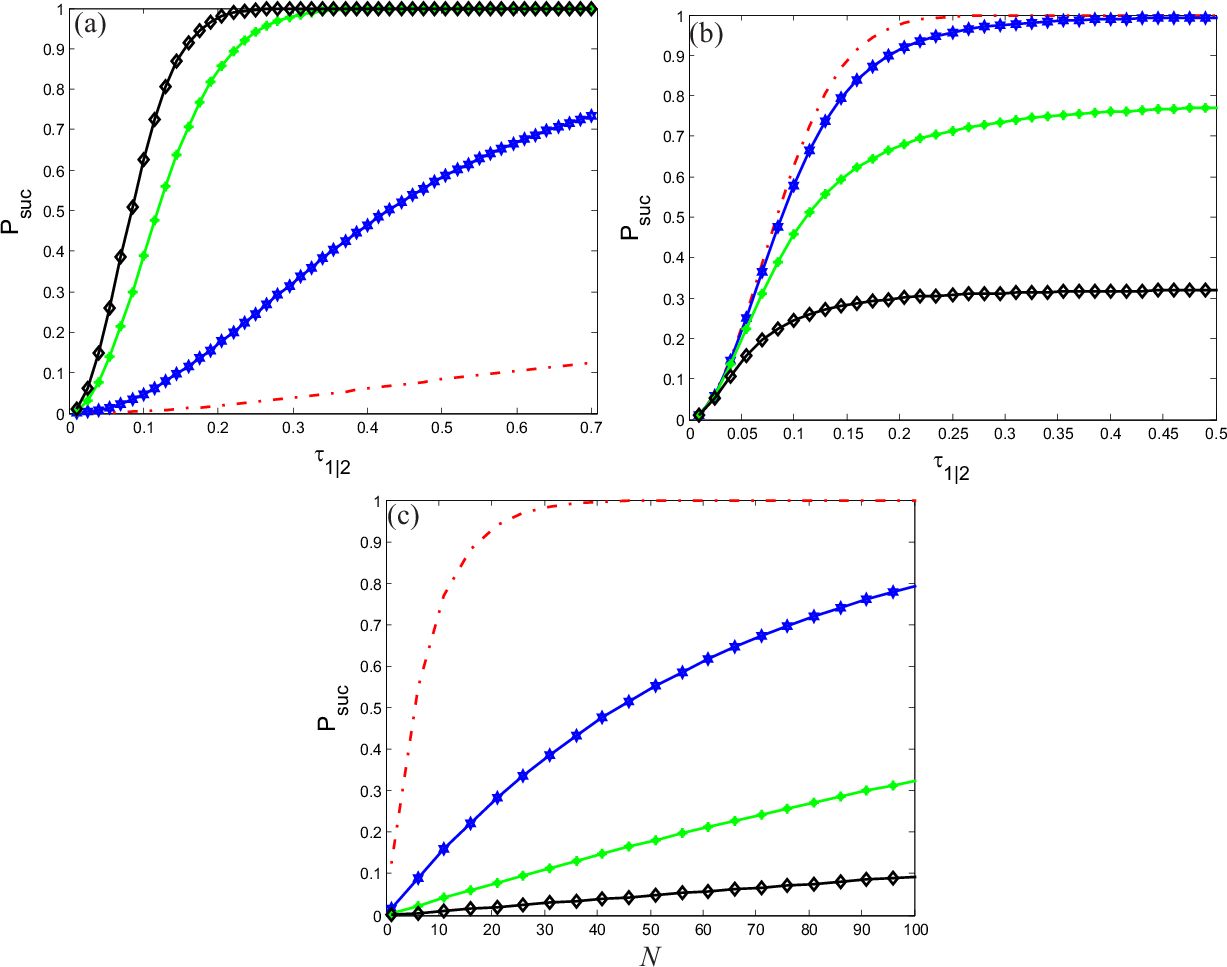}
\caption{\protect\footnotesize (a) Variation of the probability of success as a function of $\tau_{1|2}$ for various $N$.  The line marker ``$-\cdot$", ``$-$hexagram", ``$-*$" and ``$-\diamond$" represent $N=2$, $N=10$, $N=100$ and $N=200$ respectively. We choose $\tau_{2|1}=1/\sqrt{2}$. (b) Variation of the probability of success as a function of $\tau_{1|2}$ for various $\tau_{2|1}$. The line marker ``$-\cdot$", ``$-$hexagram", ``$-*$" and ``$-\diamond$" denote $\tau_{2|1}=1/\sqrt{2}$, $\tau_{2|1}=1/4$, $\tau_{2|1}=1/8$ and $\tau_{2|1}=1/16$ respectively. We consider $N=200$. (c)  Variation of the success probability as a function of nodes number. The line marker ``$-\cdot$", ``$-$hexagram", ``$-*$" and ``$-\diamond$" denote $\tau_{1\cdot2}=1/\sqrt{2}$, $\tau_{1\cdot2}=1/4$, $\tau_{1\cdot2}=1/8$ and $\tau_{1\cdot2}=1/16$ respectively. We take $\varrho=1$ in our numerical computations.} 
\label{fig3}
\end{figure*}
where $n_i=n$, and $i=1,...N$. We have denoted the four Bell states as $\left|\Phi^{\pm}\right\rangle=2^{-1/2}(\left|00\right\rangle\pm\left|11\right\rangle)$ and $\left|\Psi^{\pm}\right\rangle=2^{-1/2}(\left|01\right\rangle\pm\left|10\right\rangle)$. The $\pm^{(\varrho)}, \mp^{(\varrho)}$ and $\pm^{(\zeta)},\mp^{(\zeta)}$  represent the results corresponding to BSM on qubit pairs ($I_3,I_4$) and ($I_1,I_2$) respectively. Now, with the application of proper Pauli operator on qubit $H_4$, the entangled state of $H_4,J_1,J_2,J_3$ can be realized. For instance, if the entangled state of $H_4,J_1,J_2,J_3$ is $\left|0100\right\rangle-\left|0111\right\rangle-\left|1000\right\rangle-\left|1011\right\rangle$, applying the Pauli $z$ matrix and then the Pauli $x$, entangled state $\left|\mathcal{CS}\right\rangle$ can be realized. Now, edge route node I sends the result of the measurements along with route reply to node A through node B. Once node A receives the information, quantum channel between nodes A and J would be established. Thus the quantum state can then be transferred from node A to node J.
 
\section{Quantum wireless multihop teleportation}
In this section, we propose a quantum wireless multihop teleportation for wireless mesh backbone networks. Firstly,  let us consider the case $N=2$. From Figs. \ref{fig2} (a) and (b), it can be seen that A is a neighborhood node of J and consequently, we can infer that particles $A_3$, $J_3$, $J_1$ and $J_2$ are entangled. Now, performing Bell state measurements on particles pairs $(A_1,A_3)$ and $(A_2,J_2)$ and we obtain.
\begin{eqnarray}
\left|\Omega\right\rangle&=&\frac{1}{4}\sum_{\mathcal{X},\mathcal{Y}}\ _{A_2J_2}\left\langle\mathcal{X}^{\pm}\right| _{A_1A_3}\left\langle\mathcal{Y}^{\pm}\right.\left|\Pi'\right\rangle,\ \ \ \mathcal{X},\mathcal{Y}\in\left[\Phi,\Psi\right],\nonumber\\
&&\mbox{with}\ \ \ \left|\Pi'\right\rangle=\left|\chi\right\rangle\otimes\left|\mathcal{CS}\right\rangle
\end{eqnarray}
where we have used the following formulas for mathematical simplicity:
\begin{widetext}
\begin{eqnarray}
_{A_2J_2}\left\langle\Phi^{\pm}\right|\ _{A_1A_3}\left\langle\Phi^{\pm}\right.\left|\Pi'\right\rangle&=&\pm^{(\varsigma)}\mp^{(\sigma)}d_0\left|11\right\rangle_{J_1J_3}+a_0\left|00\right\rangle_{J_1J_3}\nonumber\\
_{A_2J_2}\left\langle\Phi^{\pm}\right|\ _{A_1A_3}\left\langle\Psi^{\pm}\right.\left|\Pi'\right\rangle&=&\pm^{(\varsigma)}\pm^{(\sigma)}d_0\left|01\right\rangle_{J_1J_3}+a_0\left|10\right\rangle_{J_1J_3}\nonumber\\
_{A_2J_2}\left\langle\Psi^{\pm}\right|\ _{A_1A_3}\left\langle\Phi^{\pm}\right.\left|\Pi'\right\rangle&=&\pm^{(\varsigma)}\pm^{(\sigma)}d_0\left|10\right\rangle_{J_1J_3}+a_0\left|01\right\rangle_{J_1J_3}\nonumber\\
_{A_2J_2}\left\langle\Psi^{\pm}\right|\ _{A_1A_3}\left\langle\Psi^{\pm}\right.\left|\Pi'\right\rangle&=&\pm^{(\varsigma)}\mp^{(\sigma)}d_0\left|00\right\rangle_{J_1J_3}-a_0\left|11\right\rangle_{J_1J_3}.
\end{eqnarray}
\end{widetext}
The $\pm^{(\varsigma)}$ and $\pm^{(\sigma)},\mp^{(\sigma)}$  represent the results corresponding to BSM on qubit pairs ($A_1,A_3$) and ($A_2,J_2$) respectively. Now node A transmits this result to node J via the selected route. By using an appropriate unitary transformation, the state $\left|\chi\right\rangle_{A_1A_2}$ can be retrieved at destination node J. Thus, the quantum communication is completed successfully. Now, let us assume that the information is transmitted to node $\mathcal{N}^{k}$, where $k=2,4,6,...2N$ denotes the number of Bell state measurement needed to be performed. Now, assuming that the state of node $\mathcal{N}^{2k}$ and J is $\left|\textbf{CS}'\right\rangle=\tau_0\left|0000\right\rangle+\tau_1\left|0011\right\rangle+\tau_2\left|1100\right\rangle-\tau_3\left|1111\right\rangle$, then we perform Bell state measurement at node A which is equal to $\mathcal{N}^{2N}$. This process has been depicted in Fig. \ref{fig2}(b). The quantum entanglement channel is represented by the dotted lines. Thus, we obtain the states of A and J as follows:
\begin{eqnarray}
\left|\Omega\right\rangle&=&\frac{1}{2}\sum_{\mathcal{A},\mathcal{B}}\ _{A_2J_2}\left\langle\mathcal{A}^{\pm}\right| _{A_1A_3}\left\langle\mathcal{B}^{\pm}\right.\left|\Pi''\right\rangle,\ \ \ \ \ \mathcal{A},\mathcal{B}\in\left[\Phi,\Psi\right],\nonumber\\
&&\mbox{with}\ \ \ \left|\Pi''\right\rangle=\left|\chi\right\rangle\otimes\left|\mathcal{CS'}\right\rangle
\end{eqnarray}
where we have used the following formulas for mathematical simplicity:
\begin{widetext}
\begin{eqnarray}\\
_{A_2J_2}\left\langle\Psi^{\pm}\right|\ _{A_1A_3}\left\langle\Psi^{\pm}\right.\left|\Pi''\right\rangle&=&\pm^{(\varsigma)}\mp^{(\sigma)}d_0\tau_0\left|00\right\rangle_{J_1J_3}-a_0\tau_3\left|11\right\rangle_{J_1J_3}
\nonumber\\
_{A_2J_2}\left\langle\Phi^{\pm}\right|\ _{A_1A_3}\left\langle\Psi^{\pm}\right.\left|\Pi''\right\rangle&=&\pm^{(\varsigma)}\pm^{(\sigma)}d_0\tau_1\left|01\right\rangle_{J_1J_3}+a_0\tau_2\left|10\right\rangle_{J_1J_3}\nonumber\\
_{A_2J_2}\left\langle\Psi^{\pm}\right|\ _{A_1A_3}\left\langle\Phi^{\pm}\right.\left|\Pi''\right\rangle&=&\pm^{(\varsigma)}\pm^{(\sigma)}d_0\tau_2\left|10\right\rangle_{J_1J_3}+a_0\tau_1\left|01\right\rangle_{J_1J_3}\nonumber\\
_{A_2J_2}\left\langle\Phi^{\pm}\right|\ _{A_1A_3}\left\langle\Phi^{\pm}\right.\left|\Pi''\right\rangle&=&\pm^{(\varsigma)}\mp^{(\sigma)}d_0\tau_3\left|11\right\rangle_{J_1J_3}+a_0\tau_0\left|00\right\rangle_{J_1J_3}.
\end{eqnarray}
\end{widetext}
Since the states of $J_1,J_3$ are not normalized, it then implies that each outcome of the measurement has different probability. In order to avoid redundancy, we shall not discuss all the outcomes but one. For other cases, node J  can apply similar approach to reconstruct the original state. Now, suppose the result of Bell state measurement is $\left|\Psi^{+}\right\rangle_{A_1,A_3}\left|\Phi^{-}\right\rangle_{A_2,J_2}$, consequently, without loss of generality, the state of qubit pair $(J_1,J_3)$ collapses to $\mathcal{G}=2^{-1}(\left.a_0\tau_2|10\right\rangle_{J_1J_3}-\left.d_0\tau_1|01\right\rangle_{J_1,J_3})$. With this result, the state $\left|\chi\right\rangle$ can be recovered at Node J. In order for this to be achieved, it is required to apply positive-operator valued measure (POVM) \cite{MA9}.

In utilizing POVM, first an ancilla (i.e., auxillary quantum system) is prepared in a known state, say $\rho_{anc}$. Combining this ancilla with the original quantum state gives an uncorrelated state. Now, the combined Hilbert space will be subjected to a maximal test which is represented by orthogonal resolution of the identity. The results of this test is related to orthogonal projector which satisfies the relations: $\mathcal{K}_\mu \mathcal{K}_\nu=\delta_{\mu\nu} \mathcal{K}_\nu$ and $\sum_\mu \mathcal{K}_\mu =1$. The probability that preparation $k$ will be followed by outcome $\mu$ is represented by $\mathcal{K}_{\mu k}=Tr(A_\mu\rho_k)$, where $A_\mu$ denotes an operator acting on the Hilbert space. The set of $A_\mu$ is called POVM \cite{MA9}. Several studies and applications of POVM have been noted down in many literature \cite{MA91,MA92,MA93,MA94}. The current study will also utilize it.

To do that, we need to set up a close indistinguishability such that the coefficient of $\left|00\right\rangle_{J_1,J_3}$ is $a_0$ and that of $\left|11\right\rangle_{J_1,J_3}$ should be $d_0$. To accomplish this, node J performs a local unitary operation $\mathcal{UT}=\sigma_x\otimes I_{2\times 2}$ on $\mathcal{G}$ to obtain $\mathcal{G}_0=2^{-1}(\left.a_0\tau_2|00\right\rangle_{J_1J_3}-d_0\tau_1\left|11\right\rangle_{J_1J_3})$.  Now, the node introduces auxiliary qubits, say $\mathcal{DE}$ with state $\left|00\right\rangle_{\mathcal{DE}}$. Entangling these qubits with $\mathcal{G}_0$ gives $\mathcal{G}_1=2^{-1}(a_0\tau_2\left|0000\right\rangle_{J_1J_3\mathcal{DE}}-d_0\tau_1\left|1100\right\rangle_{J_1J_3\mathcal{DE}})$. The node then performs a C-NOT operation on qubit pairs $(J_1,\mathcal{D})$ and $(J_3,\mathcal{E})$ to obtain 
$\mathcal{G}_2=1/4\left(a_0\left|00\right\rangle_{J_1J_3}+d_0\left|11\right\rangle_{J_1J_3}\right)\otimes\left(\tau\left|00\right\rangle_{\mathcal{DE}}-\left|11\right\rangle_{\mathcal{DE}}\right)+\left(a_0\left|00\right\rangle_{J_1J_3}-d_0\left|11\right\rangle_{J_1J_3}\right)\otimes\left(\tau_2\left|00\right\rangle_{\mathcal{DE}}+\tau_1\left|11\right\rangle_{\mathcal{DE}}\right)$. With the condition that $\tau_2\left|00\right\rangle_{\mathcal{DE}}\mp\tau_1\left|11\right\rangle_{\mathcal{DE}}$ can be conclusively discerned using a suitable measurement, state $\left(a_0\left|00\right\rangle_{J_1J_3}\pm d_0\left|11\right\rangle_{J_1J_3}\right)$ can be obtained at node J. The optimal POVM required can be written in the following subspace
\begin{eqnarray}
&&\mathcal{P}_1=\frac{1}{{\varrho}}\left|\Lambda_1\right\rangle\left\langle\Lambda_1\right|,\ \ \ \mathcal{P}_2=\frac{1}{{\varrho}}\left|\Lambda_2\right\rangle\left\langle\Lambda_2\right|,\nonumber\\
&&\mathcal{P}_3=I-\frac{1}{\varrho}\sum_{i=1}^2\left|\Lambda_i\right\rangle\left\langle\Lambda_i\right|\label{EQ4},
\end{eqnarray}
where
\begin{eqnarray}
&&\left|\Lambda_1\right\rangle=\frac{1}{\sqrt{\gamma}}\left(\frac{1}{\tau_2}\left|00\right\rangle-\frac{1}{\tau_1}\left|11\right\rangle\right)_{\mathcal{DE}},\nonumber\\
&&\left|\Lambda_2\right\rangle=\frac{1}{\sqrt{\gamma}}\left(\frac{1}{\tau_2}\left.|00\right\rangle+\frac{1}{\tau_1}\left|11\right\rangle\right)_{\mathcal{DE}},\nonumber\\
&&\mbox{with}\ \ \gamma=\frac{1}{\tau_1^2}+\frac{1}{\tau_2^2}.\label{EQ5}
\end{eqnarray}
$I$ denotes an identity operator and $\varrho$ is a parameter which defines the range of positivity of operator $\mathcal{P}_3$. Now, if the node's POVM result yields $\mathcal{P}_1$ whose probability is $\left\langle\mathcal{G}_1\right|\mathcal{P}_1\left|\mathcal{G}_1\right\rangle=1/(4\varrho\gamma)$, then one can infer the state of qubits $J_1J_3$ to be $a_0\left|00\right\rangle-d_0\left|11\right\rangle$. Afterward, the node  performs unitary operation $I_{2\times 2}\otimes\sigma_z$ on the particles in order to retrieve the original state. However, suppose the result is $\mathcal{P}_2$, with a probability calculated by $\left\langle\mathcal{G}_1\right.|\mathcal{P}_2|\left.\mathcal{G}_1\right\rangle=1/(4\varrho\gamma)$, then the node finds that the state of qubits $J_1J_3$ is $\left.a_0|00\right\rangle+d_0\left|11\right\rangle$ which is the original state of the particle.  Suppose the result is $\mathcal{P}_3$, the teleportation fails because of the node's ineptitude to infer anything about the identity of the particles state. Thus, we calculate the total probability of success as 
\begin{equation}
\text{P}_{\text{suc}}=1-\left(1-\frac{1}{2\varrho\gamma}\right)^{N-1}.\label{EQ6}
\end{equation}
Fig. \ref{fig3} (a) gives the variation of success chance as a function of parameter $\tau_{1|2}$. Taking $N=200$, as $\tau_{1|2}$ increases, success probability proliferates until it reaches $\approx1$ (at $\tau_{1|2}\approx0.25$) where no significant variation can be discerned. The same characteristic is also observed for $N=100$ except that P$_{\text{suc}}$ becomes $1$ at $\tau_{1|2}\approx0.4$. Reducing $N$ further to $10$ or $2$, we observe that the probability of achieving success teleportation dwindles. Furthermore, as anticipated, Fig. \ref{fig3}(b) shows that optimum probability would be attained with $\tau_{2|1}=1/\sqrt{2}$. Fig. \ref{fig3}(c) corroborates this fact. Also, Fig. \ref{fig3}c shows susceptibility of P$_{\text{suc}}$ to $N$. For a particular $\tau_{1|2}$, the success probability is highly sensitive to variation in $N$.

\section{Conclusion}
To sum it up, in this paper, we propose a quantum routing protocol with multihop teleportation for wireless mesh backbone networks. The quantum channel that linked the intermediate nodes has been realized through entanglement swapping based on four-qubit cluster state. After quantum entanglement swapping, quantum link was established between the source node and the destination node and quantum states are transferred via quantum teleportation. We have shown that the quantum teleportation would be successful if the sender performs a Bell state measurement, and the receiver introduces auxiliary particles, applies positive operative value measure and then utilizes corresponding unitary transformation to recover the transmitted state. We have numerically scrutinized the success probability of transferring the quantum state. This study is another supportive evidence that justifies quantum entanglement as key resource in quantum information science and furtherance of recent studies \cite{MA2,MA3,MA10}. Particularly, in Ref. \cite{MA3}, where partially entangled GHZ state was employed as quantum channel and $35$ nodes are required to achieve success probability of about $0.5$, given that transmission range is $200$m. In the current study, with $35$ nodes, success probability of $1$ is obtainable. This result agrees with the one obtained in Ref. \cite{MA3} by taking the transmission range as $300$m and consider maximally entangled GHZ states.

\end{document}